\documentclass[11pt,twoside]{article}
\usepackage{latexsym}
\usepackage{amssymb}
\usepackage{a4wide}

\usepackage[active]{srcltx}

\makeatletter

\@addtoreset{equation}{section} \makeatother
\newtheorem{theorem}{Theorem}[section]
\newtheorem{lemma}[theorem]{Lemma}
\newtheorem{proposition}[theorem]{Proposition}
\newtheorem{corollary}[theorem]{Corollary}
\newtheorem{definition}[theorem]{Definition}
\newtheorem{remark}[theorem]{Remark}

\newcommand{\alfadt}{\dot{\tilde{\alpha}}_2}

\newcommand{\cvd}{\hfill \rule{0.5em}{0.5em}}

\newcommand{\R}{{\mathbb R}}

\newtheorem{ejemplo}{Example}

\newcommand{\nn}{\mathbf{n}}
\newcommand{\vv}{\mathbf{v}}

\newcommand{\raiz}{\sqrt{\delta\Delta}}

\newcommand{\ptres}[2]{\bar{g}_3( #1,#2)}

\newcounter{cambio}

\begin{document}
\title{{\bf\Large New Examples of Marginally Trapped Surfaces and Tubes in Warped Spacetimes}}
\author{{\bf\large J.L. Flores$^*$, S. Haesen$^{\dagger}$, M. Ortega$^\ddagger$}\\
\\
{\it\small $^*$Departamento de \'Algebra, Geometr\'{\i}a y Topolog\'{\i}a,}\\
{\it \small Facultad de Ciencias, Universidad de M\'alaga,}\\
{\it\small Campus Teatinos, 29071 M\'alaga, Spain.}\\
{\it\small $^\dagger$Simon Stevin Institute for Geometry,}\\
{\it\small Mina Krusemanweg 1, 5032 ME Tilburg, The Netherlands.}\\
{\it\small $^\ddagger$Departamento de Geometr\'{\i}a y Topolog\'{\i}a,}\\
{\it\small Facultad de Ciencias, Universidad de Granada,}\\
{\it\small Avenida Fuentenueva s/n, 18071 Granada, Spain.}}

\date{}

\maketitle

\smallskip

\begin{abstract}
In the present paper we provide new examples of marginally trapped
surfaces and tubes in FLRW spacetimes by using a basic relation
between these objects and CMC surfaces in $3$-manifolds. We also
provide a new method to construct marginally trapped surfaces in
closed FLRW spacetimes, which is based on the classical Hopf map.
The utility of this method is illustrated by providing marginally
trapped surfaces crossing expanding and collapsing regions of a
closed FLRW spacetime. The approach introduced in this paper is
also extended to twisted spaces.
\end{abstract}


\noindent {\bf\footnotesize PACS:} {\scriptsize 02.40-k 04.20-q}.\\
{\it\footnotesize Key words}. {\scriptsize Warped spacetimes,
trapped surface, marginally trapped surface/tube, CMC surface,
minimal surface, Hopf fibration, Clifford tori, twisted space.}


\section{Introduction}

The notion of {\em trapped surface} was firstly introduced by
Penrose \cite{Pe'} in order to study singularities in General
Relativity. These surfaces, and their various relatives, have been
extensively studied in recent years (just to mention a few, see
e.g. \cite{BM,Bray-HuIl,Da,Leh}), since they are central not only
for singularity theorems, but also to understand the evolution of
black holes, the cosmic censorship hypothesis, the Penrose
inequality...

Trapped surfaces have the physical property that the two null
congruences normal to the surface are both converging. From the
mathematical point of view, the null converging condition means
that the mean curvature vector, which measures the tension of the
surface coming from the surrounding space, is a timelike vector
everywhere on the surface. If, in addition, the mean curvature
vector is future- or past-pointing all over the surface, the
trapped surface is accordingly called {\em future-} or {\em
past-trapped}.

In this paper, a \textit{closed surface} is a compact surface
without boundary, embedded in some other semi-Riemannian manifold.
The existence of closed trapped surfaces has been investigated in
several types of spacetimes. For example, the formation of these
surfaces in several cosmological spacetimes have been studied in
\cite{Ellis, MaMur}.

A spacelike surface in a 4-dimensional Lorentzian manifold is
called {\em marginally trapped}\footnote{This definition may
appear slightly modified in the bibliography.} if its mean
curvature vector is null. When its mean curvature vector is zero
all over the surface it is called {\em extremal}.


In order to gain some idea of the properties of marginally trapped
surfaces in particular spacetimes, classification results were
obtained for the case of having positive relative nullity in
Lorentzian space forms \cite{ChVe} and in Robertson-Walker spaces
\cite{ChVe'}. In \cite{HaOr, HaOr2, HaOr3} marginally trapped
surfaces invariant under symmetries of 4-dimensional Minkowski
space were studied. A complete classification of spacelike
surfaces in a 4-dimensional Lorentzian spacetime, containing the
above cases, was recently given in \cite{Se'}.

Some results concerning the non-existence of closed marginally
trapped surfaces can be also found in the literature. Among the
classical ones, a result due to R. Penrose \cite{Pe'} implies the
non-existence of closed marginally trapped surfaces in the
Minkowski spacetime when it bounds a compact domain. In
\cite{MaSe} the non-existence of closed marginally trapped
surfaces is shown for strictly stationary spacetimes. Finally, in
\cite{CM} the authors have shown the non-existence of marginally
trapped surfaces bounding a domain and entering a region of a
static spacetime where the Killing vector field is timelike, and
with the additional assumptions of dominant energy condition and
an outer untrapped barrier.

The main aim of this paper is to provide new examples of
marginally trapped surfaces and tubes in warped spacetimes.
\begin{enumerate}
 \item In Section \ref{s1}, we
establish some existence/non-existence results on trapped and
marginally trapped surfaces in FLRW spacetimes (Corollaries
\ref{c1}, \ref{c2}) by using a simple, but fundamental, relation
between these surfaces and constant mean curvature surfaces in
$3$-manifolds (Theorem \ref{th1}). In particular, we show the
existence of closed marginally trapped surfaces with any genus in
closed FLRW spacetimes. \item In Section \ref{s2}, we develop a
method to construct marginally trapped surfaces in closed FLRW
spacetimes, which is based on an extension of the classical Hopf
map to a submersion between closed FLRW spacetimes of dimension 4
and 3.
We illustrate it with a simple example in Subsection \ref{ss2.4}.
In order to show the utility of this constructive method, in
Subsection \ref{ss2.5} we apply it to provide marginally trapped
surfaces crossing expanding and collapsing regions of a closed
FLRW spacetime. \item Section \ref{s3} is devoted to studying {\it
marginally trapped tubes}. They are defined as
smooth hypersurfaces foliated by marginally trapped surfaces.
Then, we give some existence/non-existence results for these
objects in closed FLRW spacetimes (Corollaries \ref{c3},
\ref{c4}), and provide examples of them with any type of causal
behavior, Subsection \ref{ss3.2}. Finally, in Section \ref{s4} we
extend the approach introduced in this paper to twisted spaces
(Theorem \ref{th1'}).
\end{enumerate}

\section{Marginally trapped surfaces in $t$-slices of warped spacetimes}\label{s1}

In general, given an immersion $\mathbf{x}:\bar\Sigma^n\rightarrow
\Sigma^m$ of a $n$-dimensional semi-Riemannian manifold into
another $m$-dimensional semi-Riemannian manifold, the second
fundamental form and the mean curvature vector will be denoted by
$h_{\mathbf{x}}$ and $\vec
H_{\mathbf{x}}=\mathrm{trace}(h_{\mathbf{x}})/n$, respectively.

Let $f:I\subset{\mathbb R}\rightarrow (0,\infty)$ be a smooth
function ($t\in I$), $(M^{3},g_{3})$ a $3$-dimensional Riemannian
manifold and $S$ a surface. Denote by
$\overline{M}^{4}_{1}=I\times_{f}M^{3}$ the Lorentzian warped
product manifold given by the product manifold $I\times M^{3}$
endowed with metric $\overline{g}_{4}=-dt^{2}+f^{2}g_{3}$. Let
$\varphi:S\rightarrow M^{3}$ be an immersion of $S$ in $M^{3}$,
$\psi:M^{3}\rightarrow I\times_{f}M^{3}$ the embedding of $M^{3}$
in $I\times_{f}M^{3}$ and $\phi:S\rightarrow I\times_{f}M^{3}$ the
corresponding immersion of $S$ in the warped product, both in the
$t$ slice ($t$-{\em slice} for short). According to a well-known
result (see \cite[p. 79]{Chen}), the following relation holds
between the corresponding second fundamental forms:
\begin{equation}\label{e0}
h_{\phi}(X,Y)=h_{\varphi}(X,Y)+h_{\psi}(X,Y),\quad\hbox{whereby}\;\;
X,Y\in\mathfrak{X}(S).
\end{equation}
The expression of $h_{\psi}$ is also known (see e.g. \cite[p.
344]{ON83}):
$$h_{\psi}(X,Y)=-\overline{g}_{4}(X,Y)\frac{{\rm
grad}_{\overline{g}_{4}}f}{f}=\overline{g}_{4}(X,Y)\frac{f'}{f}\partial_{t},\quad\hbox{where
we have used}\;\; {\rm
grad}_{\overline{g}_{4}}f=-f'\partial_{t}.$$ Hence, we obtain
$h_{\phi}(X,Y)=h_{\varphi}(X,Y)+\overline{g}_{4}(X,Y)\frac{f'}{f}\partial_{t}.$
Taking one half of the
trace of the above expression, using an orthonormal frame
$\{\partial_{t}, \{E_{i}\}_{i=1}^{3}\}$ w.r.t. the metric
$\overline{g}_{4}$, i.e. $E_{i}=\frac{e_{i}}{f}$ whereby
$\{e_{i}\}_{i=1}^{3}$ is the corresponding orthonormal frame
w.r.t. the metric $g_{3}$ on $M^{3}$ (and $S$), one obtains
\begin{equation}\label{e1}
\vec{H}_{\phi}=\frac{\vec{H}_{\varphi}}{f^{2}}+\frac{f'}{f}\partial_{t},
\end{equation}
where $\vec{H}_{\phi}$ and $\vec{H}_{\varphi}$ stand for the mean
curvature vectors associated with $h_{\phi}$ and $h_{\varphi}$,
respectively.

Recall that a surface $S$ is called of {\it constant mean
curvature}, {\em CMC} for short, if the length of its mean
curvature vector is a constant function.

\begin{theorem}\label{th1} A surface $S$ contained in a $t_{0}$-slice
of $\overline{M}_{1}^{4}=I\times_{f}M^{3}$ is trapped (respectively, marginally trapped)
iff it is a CMC surface in $M^{3}$ with
\[ \|\vec{H}_{\varphi}\|<|f'(t_{0})| \quad (\mbox{respectively,}\
\|\vec{H}_{\varphi}\|=|f'(t_{0})|).
\]
\end{theorem}
{\it Proof.} We compute the length of the mean curvature $\vec H_{\phi}$, by making use of (\ref{e1}):
\[
f(t_0)^2\bar{g}(\vec H_{\phi},\vec H_{\phi}) =
f(t_0)^2 \left\|\frac{\vec{H}_{\varphi}}{f(t_{0})^{2}}+\frac{f'(t_{0})}{f(t_{0})}\partial_{t}\right\|^{2}
=\|\vec{H}_{\varphi}\|^{2} - {f'(t_{0})}^{2}.
\]
This readily gives the results.\cvd
%

\vspace{1mm}

From this result one can deduce some simple consequences for {\em
FLRW spacetimes}, i.e. warped spacetimes with fiber
$M^{3}={\mathbb R}^{3},\; {\mathbb S}^{3}\;\hbox{or}\; {\mathbb
H}^{3}$. When the fiber is $M^3=\mathbb{S}^3$, we will say that
our FLRW is \textit{closed}.

First, recall that the so-called Clifford tori $C_{u}$ in
$\mathbb{S}^3$ are given by
\[
C_{u}:=\left\{(z_{1},z_{2})\in {\mathbb C}^{2}: |z_{1}|=\cos(u),\;
|z_{2}|=\sin(u) \right\},\qquad u\in (0,\pi/2).
\]
These are closed surfaces in ${\mathbb S}^{3}$ with
constant mean curvature satisfying
\begin{equation}\label{ec:cm-toro-clifford}
\|\vec{H}_{u}\|=|2\cot(2u)|.
\end{equation}
Of course, other CMC tori can be obtained by applying to them
isometries of ${\mathbb S}^{3}$. In addition, by making surgery on
a finite number of Clifford tori, Butscher-Packard \cite{BuPa}
obtained closed surfaces in $\mathbb{S}^3$ that are also CMC with
arbitrary genus. Moreover, as far as we know, these are the only
known surfaces in $\mathbb{S}^3$ which are closed, CMC,
non-minimal and with arbitrary genus.



\begin{corollary}\label{c1} (Existence result).
There exist closed trapped and closed mar\-gi\-nally trapped
surfaces with arbitrary genus in closed FLRW spacetimes.
\end{corollary}
{\it Proof.} From Theorem \ref{th1}, any surface in $M^{3}$ with
constant mean curvature $\|\vec{H}_{\varphi}\|=|f'(t_{0})|$ can be
seen as a marginally trapped surface in the $t_{0}$-slice of
$I\times_{f}M^{3}$. Notice also that there exist closed CMC
surfaces in $M^{3}={\mathbb S}^{3}$ with arbitrary genus (standard
spheres; Clifford tori; Butscher-Packard surfaces \cite{BuPa}).
Therefore, there exist closed marginally trapped surfaces in
$I\times_{f}{\mathbb S}^{3}$ with arbitrary
genus.
From Theorem \ref{th1}, any minimal surface $S$ in $M^{3}$ can be
seen as a trapped surface in any $t_{0}$-slice of
$I\times_{f}M^{3}$ with $f'(t_{0})\neq 0$. Notice also that there
exist closed minimal surfaces in $M^{3}={\mathbb S}^{3}$ with
arbitrary genus \cite{La}. Therefore, there exist closed trapped
surfaces in $I\times_{f}{\mathbb S}^{3}$ with arbitrary genus
whenever $f\not\equiv cte$. \cvd
\begin{remark}
{\em Standard spheres and embedded CMC tori can be chosen
with any constant value of its mean curvature function, and so, there is no restriction for the warping
function $f$ in Corollary \ref{c1}. However, the mean curvature function of a Butscher-Packard's
surface has to be sufficiently small, due to the gluing process.
Thus, in this case Corollary \ref{c1} only applies to warping functions
of sufficiently small derivative. This observation must be also
taken into account in Corollary \ref{c3}.}
\end{remark}



\begin{corollary}\label{c2} (Non-existence result). Let $\overline{M}_{1}^{4}=I\times_{f} M^{3}$ be a FLRW
spacetime with fiber $M^{3}={\mathbb H}^{3}$. There are no closed
marginally trapped surfaces contained in any $t_{0}$-slice such
that $|f'(t_{0})|\leq 1$.
\end{corollary}
{\it Proof.} According to a result by do Carmo and Lawson
\cite{CaLa}, if $S$ is a closed CMC surface in ${\mathbb H}^{3}$,
it must be a geodesic sphere with mean curvature satisfying
$\|\vec{H}_{\varphi}\|>1$. Therefore, the proof directly follows
from Theorem \ref{th1}. \cvd


\begin{remark}{\em Formula (\ref{e1}) implies that $\vec{H}_{\phi}$
cannot be future-directed at $t=t_{0}$ if $f'(t_{0})\leq 0$, and
so, the following result \cite{Sen} is reobtained: {\em there are
no future trapped (resp. marginally trapped) surfaces in any slice
of collapsing (i.e. $f'(t)\leq 0$ for all $t$) warped spacetimes}.
Analogously, $\vec{H}_{\phi}$ cannot be past-directed at $t=t_{0}$
if $f'(t_{0})\geq 0$, hence: {\em there are no past trapped (resp.
marginally trapped) surfaces in any slice of expanding (i.e.
$f'(t)\geq 0$ for all $t$) warped spacetimes.}}

\end{remark}

\begin{remark}\label{ff} {\em The stability result in \cite{CM} (which can be applied to more general surfaces than the ones
contained in a $t$-slice, assumed some additional conditions)
suggests that the marginally trapped surfaces found in this
section should be unstable.}

\end{remark}



%


%
%

\section{Marginally Trapped surfaces in closed FLRW spacetimes: \newline A constructive
method}\label{s2}

In the present section we are going to construct marginally
trapped surfaces, non-necessarily contained in a $t$-slice of
warped spacetimes, by using the classical Hopf map. The price to
pay is that we will need to restrict our ambient space to closed
FLRW spacetimes.

Very roughly, the idea is as follows. We can see closed FLRW
spacetimes $I\times_{f}{\mathbb S}^{3}$ as a semi-Riemannian
submersion over $I\times_{f} {\mathbb S}^{2}(1/2)$ such that the
lift of any curve in $I\times_{f}{\mathbb S}^{2}(1/2)$ gives rise
to a surface in $I\times_{f}{\mathbb S}^{3}$ whose geometric
properties depend on the base curve. Thus, by choosing appropriate
curves in the base, we can obtain embedded surfaces in
$I\times_{f}{\mathbb S}^{3}$ with mean curvature vector as
desired, i.e. spacelike, timelike or lightlike.

To develop our approach, first we need to recall some notions
about semi-Riemannian submersions and the Hopf map.

\subsection{Semi-Riemannian submersions}\label{ss2.1}

Let $\pi:(\mathbf{M},g_M)\rightarrow (\mathbf{B},g_B)$ be a
surjective map between semi-Riemannian manifolds. Assume that
$\pi$ has maximal rank. The {\em fibers} are $\pi^{-1}(b)$, with
$b\in {\bf B}$. A tangent vector to $\mathbf{M}$ is called {\em
vertical} (resp. {\em horizontal}) if it is tangent (resp.
orthogonal) to the fibers. The {\em vertical part} of $\pi$ at a
point $m\in {\bf M}$ is $\ker (d\pi)_{m}\subset T_{m}\mathbf{M}$.
If for each point $m\in \mathbf{M}$, $\pi_{*}$ satisfies
\begin{equation}g_M(u,v)=g_B(\pi_*u,\pi_*v),
\label{submersion}
\end{equation}
for any horizontal tangent vectors $u,v$ at $m\in \mathbf{M}$,
then $\pi$ is called a \textit{semi-Riemannian submersion}. This
lemma summarizes the basic properties of semi-Riemannian
submersions, \cite{ON}.

\begin{lemma}\label{submersion-prop} Let $\pi:(\mathbf{M},g_M)\rightarrow (\mathbf{B},g_B)$ be a
semi-Riemannian submersion.
\begin{itemize}

\item Given $X\in \mathfrak{X}({\bf B})$, there exists a {\em
horizontal lift} $\tilde{X}\in \mathfrak{X}({\bf M})$ of $X$ such
that $\tilde{X}$ is horizontal and $\pi_{*}\tilde{X}=X$

\item Given a curve $\gamma:I\rightarrow {\bf B}$, $t_{0}\in I$
and a point $m\in \pi^{-1}(\gamma(t_{0}))$, there exists a unique
{\em horizontal lift} $\tilde{\gamma}:I\rightarrow \mathbf{M}$ of
$\gamma$, i.e. it satisfies $\tilde{\gamma}(t_{0})=m$,
$\pi\circ\tilde{\gamma}=\gamma$ and $\tilde{\gamma}'$ is
horizontal. In particular, $\gamma$ is unitary if, and only if, so
is $\tilde{\gamma}$.

\item If $\nabla^M$ and $\nabla^B$ are the Levi-Civita connections
of $\mathbf{M}$ and $\mathbf{B}$, resp., then for any
$X,Y,Z\in\mathfrak{X}(\mathbf{B})$,
$g_M(\nabla^M_{\tilde{X}}\tilde{Y},\tilde{Z})=g_B(\nabla^B_XY,Z)$.
\end{itemize}

\end{lemma}

\subsection{The Hopf map and closed FLRW spacetimes}\label{ss2.2}

Let ${\mathbb C}$ be the field of complex numbers, with
$i=\sqrt{-1}$ the complex unit, $|z|$ the modulus of $z\in
{\mathbb C}$ and $\overline{z}$ its complex conjugate. Firstly,
the round $3$-sphere in ${\mathbb C}^{2}$ can be seen as ${\mathbb
S}^{3}=\{(z,w)\in {\mathbb C}^{2}: |z|^{2}+|w|^{2}=1\}$, with
standard metric $g_{3}$. Also, we can see the round $2$-sphere of
radius $1/2$ as ${\mathbb S}^{2}(1/2)=\{(z,x)\in {\mathbb C}\times
{\mathbb R}: |z|^{2}+x^{2}=1/4\}$, with standard metric $g_{2}$.
We recall the classical {\em Hopf map}
\[
\pi: {\mathbb S}^{3}\rightarrow {\mathbb S}^{2}(1/2),\quad
\pi(z,w)=\Big(z\overline{w},\frac{1}{2}|z|^{2}-\frac{1}{2}|w|^{2}\Big),
\]
where $\overline{\omega}$ is the complex conjugate of $\omega$. It
is well-known that $\pi$ is a Riemannian submersion with totally
geodesic fibers. In fact, this Riemannian submersion $\pi$ is the
quotient map of the following isometry group action:
\begin{equation}\label{eq}
{\mathbb S}^{1}\times {\mathbb S}^{3}\rightarrow {\mathbb
S}^{3},\quad (e^{i\theta},(z,w))\mapsto
(e^{i\theta}z,e^{i\theta}w).
\end{equation}
The fibers of $\pi$ are the orbits of the action, i.e. given a
point $p=(z,w)\in {\mathbb S}^{3}$, the orbit is
$\{e^{i\theta}\cdot p=(e^{i\theta}z,e^{i\theta}w): e^{i\theta}\in
{\mathbb S}^{1}\}$, which is a big circle (geodesic) of ${\mathbb
S}^{3}$. We also remark that the vertical part of $\pi$ at
$p=(z,w)\in {\mathbb S}^{3}$ is spanned by $ip=(iz,iw)$. In other
words, ${\rm ker}(d\pi)_{p}={\rm Span}\{ip\}$.

Given $f:I\subset {\mathbb R}\rightarrow (0,\infty)$ a smooth
function ($t\in I$), consider the closed FLRW spacetime
$\overline{M}^{4}_{1}=I\times_{f}{\mathbb S}^{3}$.
We also consider the {\em toy model}
$\overline{M}^{3}_{1}=I\times_{f}{\mathbb S}^{2}(1/2)$, i.e. the
$3$-dimensional Lorentzian manifold formed by the product manifold
$I\times {\mathbb S}^{2}(1/2)$ endowed with metric
$\overline{g}_{3}=-dt^{2}+f^{2}g_{2}$. Let $\overline{\nabla}$,
$D$, $\nabla$ and $\nabla^{2}$ be the Levi-Civita connections of
$\overline{M}^{4}_{1}$, ${\mathbb S}^{3}$, $\overline{M}^{3}_{1}$
and ${\mathbb S}^{2}(1/2)$, resp.
Note that the natural projection map of $\overline{M}^{4}_{1}$
onto $I$ is a semi-Riemannian submersion, whose horizontal part is
spanned by $\partial_{t}$. Then, a vertical vector is orthogonal
to $\partial_{t}$. Given a vector field $X\in\mathfrak{X}({\mathbb
S}^{3})$, there exists a {\em vertical} lift $\tilde{X}$ tangent
to $\overline{M}^{4}_{1}$ such that $\tilde{X}\perp
\partial_{t}$. Given $Z$ a tangent vector to
$\overline{M}^{4}_{1}$, ${\rm nor}(Z)$ is the orthogonal
projection onto the horizontal part, whereas ${\rm tan}(Z)$ is the
orthogonal projection onto the vertical part. Formally, there is a
similar situation for $\overline{M}^{3}_{1}$, so that we can use
the same notation. Thus, we obtain \cite{ON83}:
\begin{lemma}\label{tan-nor} Let $X$, $Y$ be tangent vector fields
to ${\mathbb S}^{3}$ (resp. ${\mathbb S}^{2}(1/2)$) and
$\tilde{X}$, $\tilde{Y}$ be vertical lifts to
$\overline{M}^{4}_{1}$ (resp. $\overline{M}^{3}_{1}$):
\begin{itemize}

\item[1.] ${\rm
nor}(\overline{\nabla}_{\tilde{X}}\tilde{Y})=-\frac{\overline{g}_{4}(\tilde{X},\tilde{Y})}{f}{\rm
grad}_{\overline{g}_{4}}(f)$ \qquad\qquad(resp. ${\rm
nor}(\nabla_{\tilde{X}}\tilde{Y})=-\frac{\overline{g}_{3}(\tilde{X},\tilde{Y})}{f}{\rm
grad}_{\overline{g}_{3}}(f)$) \item[2.] ${\rm
tan}(\overline{\nabla}_{\tilde{X}}\tilde{Y})$ is the vertical lift
of $D_{X}Y$ \quad(resp. ${\rm tan}(\nabla_{\tilde{X}}\tilde{Y})$
is the vertical lift of $\nabla^{2}_{X}Y$).
\end{itemize}

\end{lemma}

\subsection{Constructing the surface}\label{ss2.3}

From now on, we will make use of Lemma \ref{submersion-prop},
sometimes without indicating it explicitly. We define the
projection $\overline{\pi}:\overline{M}^{4}_{1}\rightarrow
\overline{M}^{3}_{1}$ as $\overline{\pi}(t,p):=(t,\pi(p))$.
\begin{lemma}\label{star} The map $\overline{\pi}$ is a semi-Riemannian
submersion with vertical part at $(t,p)$ spanned by $(0,ip)$.
\end{lemma}
{\it Proof.} We denote by $\partial_{t}$ both, the vector field
tangent to $\overline{M}^{4}_{1}$ and $\overline{M}^{3}_{1}$.
Then,
$$\overline{\pi}_{*}\partial_{t}\mid_{(t,p)}=\frac{d}{ds}\mid_{s=0}\overline{\pi}(t+s,p)=\frac{d}{ds}\mid_{s=0}(t+s,\pi(p))=\partial_{t}\mid_{(t,\pi(p))}.$$
Therefore,
$$\overline{g}_{3}(\pi_{*}\partial_{t},\pi_{*}\partial_{t})=-1=\overline{g}_{4}(\partial_{t},\partial_{t}).$$
For any $(t,p)\in \overline{M}^{4}_{1}$, consider the curve
$\alpha(s)=(t,\cos(s)p+\sin(s)ip)$. Taking into account that ${\rm
ker}(d\pi)_{p}={\rm Span}(ip)$, we deduce:
$$\overline{\pi}_{*}(0,ip)=\frac{d}{ds}\mid_{s=0}\overline{\pi}(\alpha(s))=(0,\pi_{*}(ip))=(0,0).$$
Finally, take $(0,X)\in T_{(t,p)}\overline{M}_{1}^{4}$ which is
orthogonal to $ip$. Then, $\overline{\pi}_{*}(0,X)=(0,\pi_{*}X)$.
Taking into account that $\pi$ is a semi-Riemannian submersion, we
deduce
\[
\begin{array}{r}
\overline{g}_{3}(\overline{\pi}_{*}(0,X),\overline{\pi}_{*}(0,X))=\overline{g}_{3}((0,\pi_{*}X),(0,\pi_{*}X))=f^{2}g_{2}(\pi_{*}X,\pi_{*}X)\qquad\qquad\qquad\qquad\qquad
\\
\qquad\qquad\qquad\qquad\qquad\qquad\qquad\qquad\qquad\qquad\qquad
=f^{2}g_{3}(X,X)=\overline{g}_{4}((0,X),(0,X)).\;\cvd
\end{array}
\]

\vspace{2mm}

We recall that the Hopf map $\pi$ is the quotient map of the
isometry group action (\ref{eq}). We call
$\Gamma_{\theta}:{\mathbb S}^{3}\rightarrow {\mathbb S}^{3}$,
$\Gamma_{\theta}(z,w)=(e^{i\theta}z,e^{i\theta}w)$, which is an
isometry of ${\mathbb S}^{3}$. We extend it to
$\overline{M}^{4}_{1}$ as follows. For each $e^{i\theta}\in
{\mathbb S}^{1}$, we define the map
\[
\overline{\Gamma}_{\theta}:\overline{M}^{4}_{1}\rightarrow\overline{M}^{4}_{1},\quad
\overline{\Gamma}_{\theta}(t,p)=(t,\Gamma_{\theta}(p)).
\]
Given $(t,p)\in\overline{M}^{4}_{1}$, consider $T_{t}I\equiv
{\mathbb R}$ and $T_{p}{\mathbb S}^{3}\subset {\mathbb C}^{2}$.
Thus, it is possible to let $\overline{\Gamma}_{\theta}$ act on
tangent vectors under these natural identifications.
\begin{lemma}\label{l1} For each $e^{i\theta}\in {\mathbb S}^{1}$, the map
$\overline{\Gamma}_{\theta}$ is an isometry of
$\overline{M}^{4}_{1}$ with
$(\overline{\Gamma}_{\theta})_{*}=\overline{\Gamma}_{\theta}$
under previous identifications.
\end{lemma}
{\it Proof.} Firstly, we are going to show that
$(\overline{\Gamma}_{\theta})_{*}=\overline{\Gamma}_{\theta}$.
Observe that:
$$(\overline{\Gamma}_{\theta})_{*}(\partial_{t}\mid_{(t,p)})=\frac{d}{ds}\mid_{s=0}\overline{\Gamma}_{\theta}(t+s,p)=\frac{d}{ds}\mid_{s=0}(t+s,\Gamma_{\theta}(p))=\partial_{t}\mid_{(t,e^{i\theta}p)}=\overline{\Gamma}_{\theta}(\partial_{t}\mid_{(t,p)}).$$
On the other hand, given a curve $\gamma$ in ${\mathbb S}^{3}$
such that $\gamma(0)=p$, $\dot{\gamma}(0)=X$, we have
$$(\overline{\Gamma}_{\theta})_{*}(0,X)=\frac{d}{ds}\mid_{s=0}\overline{\Gamma}_{\theta}(0,\gamma(s))=\frac{d}{ds}\mid_{s=0}(0,e^{i\theta}\gamma(s))=(0,e^{i\theta}X)=\overline{\Gamma}_{\theta}(0,X).$$
Finally, in order to prove that $\overline{\Gamma}_{\theta}$ is
isometry, notice that
\[
\begin{array}{rl}
\overline{g}_{4}((\overline{\Gamma}_{\theta})_{*}(0,X),(\overline{\Gamma}_{\theta})_{*}(0,X))
&
=\overline{g}_{4}((0,e^{i\theta}X),(0,e^{i\theta}X))=f^{2}g_{3}(e^{i\theta}X,e^{i\theta}X)
\\ & \quad\qquad\qquad\qquad\qquad =f^{2}g_{3}(X,X)=\overline{g}_{4}((0,X),(0,X)).
\quad\quad \cvd
\end{array}
\]

Since ${\mathbb S}^{2}(1/2)$ is an orientable manifold,
$\overline{M}^{3}_{1}$ is also orientable. We choose the
orientation on $\overline{M}^{3}_{1}$ in such a way that for any
local positive tangent frame $\{X,Y\}$ on ${\mathbb S}^{2}(1/2)$,
the set $\{\partial_{t},X,Y\}$ is a local positive frame on
$\overline{M}^{3}_{1}$.

Let $\alpha:J\subset {\mathbb R}\rightarrow \overline{M}^{3}_{1}$
be a unit spacelike Frenet curve with frenet apparatus
$\{T=\dot\alpha,N,B\}$ and $\kappa>0$, $\tau$. This means that the
Frenet equations are
\begin{equation}\label{sstar}
\nabla_{T}T=\epsilon_{2}\kappa N,\quad \nabla_{T}N=\kappa
T+\epsilon_{3}\tau B,\quad \nabla_{T}B=-\epsilon_{2}\tau N,
\end{equation}
where $\epsilon_{2}=\overline{g}_{3}(N,N)$,
$\epsilon_{3}=\overline{g}_{3}(B,B)$,
$\epsilon_{2}=-\epsilon_{3}=\pm 1$, and $\{T,N,B\}$ is a positive
basis along $\alpha$. Consider $\alpha(s)=(t(s),\alpha_{2}(s))$,
where $t:J\rightarrow I$, $\alpha_{2}:J\rightarrow {\mathbb
S}^{2}(1/2)$. By Lemma \ref{submersion-prop}, let $\beta: J\subset
{\mathbb R}\rightarrow\overline{M}_{1}^{4}$ be a horizontal lift
of $\alpha$. Since $\dot\beta$ is orthogonal to the vertical part
of
$\overline{\pi}$, we have 
\[\overline{\pi}\circ\beta=\alpha, \quad
\beta=(t,\beta_{2}),\quad {\pi}\circ\beta_{2}=\alpha_{2},\quad
\dot{\beta}_{2}\perp i\beta_{2}.
\]

Now, we are able to construct a spacelike surface in
$\overline{M}^{4}_{1}$ with the help of
$\overline{\Gamma}_{\theta}$ and $\beta$. Define:
\begin{equation}\label{immersion}
\phi:S=J\times{\mathbb S}^{1}\rightarrow
\overline{M}^{4}_{1},\quad
\phi(s,\theta)=\overline{\Gamma}_{\theta}(\beta(s))=(t(s),e^{i\theta}\beta_{2}(s)).
\end{equation}
It is clear that the derivatives of $\phi$ are
\[
\phi_{s}=(\dot{t},e^{i\theta}\dot{\beta}_{2}),\quad
\phi_{\theta}=(0,ie^{i\theta}\beta_{2}).
\]
By using Lemma \ref{star}, the coefficients of the first
fundamental form of $\phi^{*}\overline{g}_{4}$ are the following:
\[
\begin{array}{rl}
E &
=\overline{g}_{4}(\phi_{s},\phi_{s})=\overline{g}_{4}((\dot{t},e^{i\theta}\dot{\beta}_{2}),(\dot{t},e^{i\theta}\dot{\beta}_{2}))=-\dot{t}^{2}+f^{2}g_{3}(e^{i\theta}\dot{\beta}_{2},e^{i\theta}\dot{\beta}_{2})
\\ & =-\dot{t}^{2}+f^{2}g_{3}(\dot{\beta}_{2},\dot{\beta}_{2})=\overline{g}_{4}(\dot{\beta},\dot{\beta})=1,
\\
F &
=\overline{g}_{4}(\phi_{s},\phi_{\theta})=\overline{g}_{4}((\dot{t},e^{i\theta}\dot{\beta}_{2}),(0,ie^{i\theta}\beta_{2}))=f^{2}g_{3}(e^{i\theta}\dot{\beta}_{2},ie^{i\theta}\beta_{2})
\\ & =f^{2}g_{3}(\dot{\beta}_{2},i\beta_{2})=0,
\\
G &
=\overline{g}_{4}(\phi_{\theta},\phi_{\theta})=\overline{g}_{4}((0,ie^{i\theta}\beta_{2}),(0,ie^{i\theta}\beta_{2}))=f^{2}g_{3}(ie^{i\theta}\beta_{2},ie^{i\theta}\beta_{2})
\\ & =f^{2}g_{3}(\beta_{2},\beta_{2})=f^{2}.
\end{array}
\]
Therefore, $\{U_{1}=\phi_{s},U_{2}=(1/f)\phi_{\theta}\}$ is a
globally defined orthonormal tangent frame to $S$ in
$\overline{M}^{4}_{1}$. We also need to construct an orthonormal
normal frame. To do so, we use the isometries
$\overline{\Gamma}_{\theta}$ and the vectors $N$, $B$ along
$\alpha$. By Lemma \ref{submersion-prop}, let $\tilde{N}$ and
$\tilde{B}$ be horizontal lifts of $N$ and $B$, resp., along
$\beta$. Define
\[
\eta_{N},\eta_{B}:S\rightarrow T\overline{M}_{1}^{4},\quad
\eta_{N}=(\overline{\Gamma}_{\theta})_{*}\tilde{N},\;\;
\eta_{B}=(\overline{\Gamma}_{\theta})_{*}\tilde{B}.
\]
\begin{lemma} The set $\{\eta_{N},\eta_{B}\}$ is a globally
defined, orthonormal, normal frame to $S$.
\end{lemma}
{\it Proof.} We note that
\[
\phi_{s}=(\overline{\Gamma}_{\theta})_{*}\dot{\beta},\quad
\phi_{\theta}=(\overline{\Gamma}_{\theta})_{*}(0,i\beta_{2}).
\]
Bearing in mind these two expressions, (\ref{submersion}) and
Lemma \ref{l1}, we deduce
\[
\begin{array}{l}
\overline{g}_{4}(\eta_{N},\phi_{s})=\overline{g}_{4}((\overline{\Gamma}_{\theta})_{*}\tilde{N},(\overline{\Gamma}_{\theta})_{*}\dot{\beta})=\overline{g}_{4}(\tilde{N},\dot{\beta})=\overline{g}_{3}(N,T)=0,
\\
\overline{g}_{4}(\eta_{N},\phi_{\theta})=\overline{g}_{4}((\overline{\Gamma}_{\theta})_{*}\tilde{N},(\overline{\Gamma}_{\theta})_{*}(0,i\beta_{2}))=\overline{g}_{4}(\tilde{N},(0,i\beta_{2}))=0,
\end{array}
\]
where the last equality holds because $\tilde{N}$ is horizontal
and $(0,i\beta_{2})$ is vertical. On the other hand,
\[
\begin{array}{l}
\overline{g}_{4}(\eta_{B},\phi_{s})=\overline{g}_{4}((\overline{\Gamma}_{\theta})_{*}\tilde{B},(\overline{\Gamma}_{\theta})_{*}\dot{\beta})=\overline{g}_{4}(\tilde{B},\dot{\beta})=\overline{g}_{3}(B,T)=0
\\
\overline{g}_{4}(\eta_{B},\phi_{\theta})=\overline{g}_{4}((\overline{\Gamma}_{\theta})_{*}\tilde{B},(\overline{\Gamma}_{\theta})_{*}(0,i\beta_{2}))=\overline{g}_{4}(\tilde{B},i\dot{\beta})=0
\\
\overline{g}_{4}(\eta_{N},\eta_{N})=\overline{g}_{4}((\overline{\Gamma}_{\theta})_{*}\tilde{N},(\overline{\Gamma}_{\theta})_{*}\tilde{N})=\overline{g}_{4}(\tilde{N},\tilde{N})=\overline{g}_{3}(N,N)=\epsilon_{2}.
\end{array}
\]
Similarly, we deduce $\overline{g}_{4}(\eta_{N},\eta_{B})=0$ and
$\overline{g}_{4}(\eta_{B},\eta_{B})=\epsilon_{3}$. \cvd

\vspace{1mm}

Let $h_{\phi}$ be the second fundamental form of the inmersion
$\phi:S\rightarrow\overline{M}^{4}_{1}$, and
$\vec{H}_{\phi}=\frac{1}{2}{\rm
trace}_{\,\overline{g}_{4}}h_{\phi}$ the corresponding mean
curvature vector of $S$ in $\overline{M}^{4}_{1}$. Then:
\begin{lemma}\label{meanofphi} The mean curvature of $\phi$ is given
by
\[
\begin{array}{rl}
\vec{H}_{\phi}(s,e^{i\theta})= &
\displaystyle\frac{\epsilon_{2}}{2}\left(
\kappa(s)+\frac{f'(t(s))}{f(t(s))}\overline{g}_{3}(\partial_{t}\mid_{\alpha(s)},N(s))\right)\cdot\eta_{N}(s)
\\
&\displaystyle
+\frac{\epsilon_{3}}{2}\left(\frac{f'(t(s))}{f(t(s))}\overline{g}_{3}(\partial_{t}\mid_{\alpha(s)},B(s))\right)\cdot\eta_{B}(s).
\end{array}
\]
\end{lemma}
{\it Proof.} Since we already know $U_{1}$, $U_{2}$, $\eta_{N}$
and $\eta_{B}$, we have
\[
2\vec{H}_{\phi}=\Sigma_{i=1}^{2}\{\epsilon_{2}\overline{g}_{4}(h_{\phi}(U_{i},U_{i}),\eta_{N})\eta_{N}+\epsilon_{3}\overline{g}_{4}(h_{\phi}(U_{i},U_{i}),\eta_{B})\eta_{B}\}.
\]
We are going to compute all these four products. By Lemma
\ref{submersion-prop} and (\ref{sstar})
\[
\begin{array}{rl}
\overline{g}_{4}(h_{\phi}(U_{1},U_{1}),\eta_{N}) &
 =\overline{g}_{4}(\overline{\nabla}_{U_{1}}U_{1},\eta_{N})=\overline{g}_{4}(\overline{\nabla}_{\phi_{s}}\phi_{s},\eta_{N})
\\
& =\overline{g}_{4}(\overline{\nabla}_{(\overline{\Gamma}_{\theta})_{*}\dot\beta} (\overline{\Gamma}_{\theta})_{*}\dot\beta,(\overline{\Gamma}_{\theta})_{*}\tilde{N})
=\overline{g}_{4}(\overline{\nabla}_{\dot{\beta}}\dot{\beta},\tilde{N})
\\
& =\overline{g}_{3}(\nabla_{\dot{\alpha}}\dot{\alpha},N)=\overline{g}_{3}(\epsilon_{2}\kappa
N,N)  =\kappa.
\end{array}
\]
Similarly, we have
\[
\begin{array}{rl}
\overline{g}_{4}(h_{\phi}(U_{1},U_{1}),\eta_{B}) &
=\overline{g}_{4}(\overline{\nabla}_{U_{1}}U_{1},\eta_{B})=\overline{g}_{4}(\overline{\nabla}_{\dot{\beta}}\dot{\beta},\tilde{B})=\overline{g}_{3}(\nabla_{\dot{\alpha}}\dot{\alpha},B)
\\ &  =\overline{g}_{3}(\epsilon_{2}\kappa N,B)  =0.
\end{array}
\]
For $U_{2}$, we make use of Lemma \ref{tan-nor}. We consider the
surface in $\mathbb{S}^3$ given by
$$ \xi: J\times \mathbb{S}^1\rightarrow \mathbb{S}^3, \quad
\xi(s,e^{i\theta})=e^{i\theta}\beta_2(s).$$ Since $\xi_{\theta} =
ie^{i\theta}\beta_2(s)$, it is clear that
$\phi_{\theta}=(0,\xi_{\theta})$. Also, note that
$U_{2}=\frac{1}{f}\phi_{\theta}=\frac{1}{f}(0,\xi_{\theta})$.
Since
$h_{\phi}(U_{2},U_{2})=\frac{1}{f^{2}}h_{\phi}(\phi_{\theta},\phi_{\theta})$,
we compute
\begin{equation}\label{tannabla}
\mathrm{tan}(\overline{\nabla}_{\phi_{\theta}}\phi_{\theta})={\rm
tan}(\overline{\nabla}_{(0,\xi_{\theta})}(0,\xi_{\theta}))=(0,D_{\xi_{\theta}}\xi_{\theta}).
\end{equation}
Thus, we have to compute $D_{\xi_{\theta}}\xi_{\theta}$. To do so,
we recall that the position vector $\chi: \mathbb{S}^3\rightarrow
\mathbb{C}^2$ is a unit normal vector field with second
fundamental form $h_{\chi}(X,Y)=-g_3(X,Y)\chi$ for any $X,Y$
tangent to $\mathbb{S}^3$. Let $\bar{D}$ be the Levi-Civita
connection of $\mathbb{C}^2$. By the Gauss formula, and by the
fact that $\xi_{\theta}$ is unit,
$$ D_{\xi_{\theta}} \xi_{\theta} =
\bar{D}_{\xi_{\theta}}\xi_{\theta}-h_{\chi}(\xi_{\theta},\xi_{\theta})=\bar{D}_{\xi_{\theta}}\xi_{\theta}+\chi\circ\xi.
$$
Now, we consider the curve in $J\times \mathbb{S}^1$ given by
$\alpha(u)=(s,e^{i(\theta+u)})$. Since $\alpha(0)=(s,e^{i\theta})$
and $\alpha'(0)=\partial_{\theta}\vert_{(s,e^{i\theta})}$, we
obtain
$$\bar{D}_{\xi_{\theta}}\xi_{\theta} =
\left.\frac{d}{du}\right\vert_{u=0} \xi_{\theta}(\alpha(u)) =
\left.\frac{d}{du}\right\vert_{u=0}
\xi_{\theta}(s,e^{i(\theta+u)})
\left.=\frac{d}{du}\right\vert_{u=0} i\,e^{i(\theta+u)}\beta_2(s)
= -e^{i(\theta)}\beta_2(s).
$$
Finally, we see
$$ D_{\xi_{\theta}} \xi_{\theta} = - e^{i\theta}\beta_2(s) + (\chi\circ
\xi)(s,e^{i\theta})=0.
$$
By (\ref{tannabla}), we see that
$\mathrm{tan}(\overline{\nabla}_{\phi_{\theta}}\phi_{\theta})=0$.
On the other hand, by Lemma \ref{tan-nor}
\[
{\rm nor}(\overline{\nabla}_{\phi_{\theta}}\phi_{\theta})={\rm
nor}(\overline{\nabla}_{(0,Z)}(0,Z))=-\frac{\overline{g}_{4}((0,Z),(0,Z))}{f}{\rm
grad}_{\overline{g}_{4}}(f)=f
g_{3}(Z,Z)f'\partial_{t}=ff'\partial_{t}.
\]
Therefore, we obtain
$\overline{\nabla}_{\phi_{\theta}}\phi_{\theta}=ff'\partial_{t}$.
As a consequence, we get
\[
\begin{array}{c}
h_{\phi}(U_{2},U_{2})=\frac{\epsilon_{2}}{f^{2}}\overline{g}_{4}(\overline{\nabla}_{\phi_{\theta}}\phi_{\theta},\eta_{N})\eta_{N}+\frac{\epsilon_{3}}{f^{2}}\overline{g}_{4}(\overline{\nabla}_{\phi_{\theta}}\phi_{\theta},\eta_{B})\eta_{B}\qquad\qquad\qquad\qquad\qquad\qquad
\\
\qquad\qquad\qquad\qquad\qquad\qquad\qquad\qquad\qquad=\frac{\epsilon_{2}f'}{f}\overline{g}_{3}(\partial_{t},N)\eta_{N}+\frac{\epsilon_{3}f'}{f}\overline{g}_{3}(\partial_{t},B)\eta_{B}.
\cvd
\end{array}
\]
\begin{proposition} Given $(s,e^{i\theta})\in S$, the mean
curvature vector of the immersion $\phi$ is spacelike (resp.
lightlike/zero, timelike) if, and only if,
\begin{equation}\label{e2}
\epsilon_{2}\left(\kappa+\frac{f'}{f}\overline{g}_{3}(\partial_{t},N)\right)^{2}+\epsilon_{3}\left(\frac{f'}{f}\overline{g}_{3}(\partial_{t},B)\right)^{2}>0\quad
(\hbox{resp.}\;\; =0,\; <0).
\end{equation}
\end{proposition}
\begin{remark} {\em Let us assume that the curve $\alpha$ satisfies that for some $s_0\in J$, the point  $t_0=t(s_0)$ is such that $f'(t_0)=0$. According to (\ref{e2}), the mean curvature vector of $\phi$ is lightlike/zero (respectively, timelike or spacelike) iff
$\epsilon_2 \kappa^2(s_0)=0$ (resp. $<0$ or $>0$). Moreover, in
case the equality holds, the curvature at $s_0$ must vanish, i.e.
$\kappa(s_0)=0$, and so our procedure is no longer valid at this
point (recall that we assumed a Frenet basis with the assumption
$\kappa>0$). As we will see later, sometimes we can overcome this
difficulty by using a continuity argument.
}
\end{remark}

Finally, we point out that given a curve $\alpha_{2}:J\subset
{\mathbb R}\rightarrow {\mathbb S}^{2}(1/2)$, the lift of
$\alpha_{2}$ to ${\mathbb S}^{3}$ via the Hopf projection $\pi$ is
classically called a {\em Hopf tube}. If, in addition, $J={\mathbb
R}$, $\alpha_{2}$ is periodic and $\alpha_{2}$ has no
self-intersection points (in other words, the image of
$\alpha_{2}$ is homeomorphic to a circle), the Hopf tube is an
(embedded) Clifford torus in ${\mathbb S}^{3}$. This clearly
extends to our curves $\alpha$ and immersions $\phi$; that is, if
the image of $\alpha:{\mathbb R}\rightarrow \overline{M}^{3}_{1}$
is homeomorphic to a circle, the associated lift $\phi$ is a torus
without boundary embedded in $\overline{M}^{4}_{1}$.

\subsection{A simple example}\label{ss2.4}

Let $\gamma:\tilde{J}\subset\R\rightarrow\mathbb{S}^2(1/2)$ be a
Frenet unit curve. Let $\nabla^2$ be the Levi-Civita connection of
$\mathbb{S}^2(1/2)$. The Frenet apparatus of $\gamma$ is
$\{d\gamma/d\tilde{s},\nn\}$ with geodesic curvature $ c $. In
other words,
$$ \nabla^2_{d\gamma/d\tilde{s}}d\gamma/d\tilde{s} =  c  \nn, \qquad \nabla^2_{d\gamma/d\tilde{s}}\nn = - c\cdot d\gamma/d\tilde{s},
$$
and $\{d\gamma/d\tilde{s},\nn\}$ is a positive basis for the usual
orientation on $\mathbb{S}^2(1/2)$. Given $t_0\in I$, let us
define $\alpha_2:J\subset\R\rightarrow\mathbb{S}^2(1/2)$,
$\alpha_2(s)=\gamma(s/f(t_0))$, and the curve
$\alpha:J\rightarrow\overline{M}_1^3$,
$\alpha(s)=(t_0,\alpha_2(s))$. Simple computations give
$\dot{\alpha}=(0,\dot{\alpha_2})=\left(0,\frac{1}{f}\frac{d\gamma}{d\tilde{s}}\right)$.
We call $\vv=\frac{1}{f}\frac{d\gamma}{d\tilde{s}}$. Note that
$(0,\vv)$ is unit for $\bar{g}_3$, and therefore, $\alpha$ is a
spacelike unit curve in $\overline{M}_1^3$.
Then, we easily obtain
$$  \nabla^2_{\vv}\vv = \frac{ c }{f^2} \nn, \qquad \nabla^2_{\vv}\nn = - c  \vv.
$$
We put $T=\dot{\alpha}=(0,\vv)$. From Lemma \ref{tan-nor}, we have
$\nabla_TT=\nabla_{(0,\vv)}(0,\vv)=
\mathrm{tan}\Big(\nabla_{(0,\vv)}(0,\vv)\Big)+
\mathrm{nor}\Big(\nabla_{(0,\vv)}(0,\vv)\Big)=
\Big(0,\nabla^2_{\vv}\vv\Big)-\frac{\ptres{(0,\vv)}{(0,\vv)}
}{f}\mathrm{grad}_{\overline{g}_{3}}f = \Big (\frac{f'}{f},\frac{
c }{f^2}\nn\Big)$. The square $\overline{g}_{3}$-norm of
$\nabla_{T}T$ is
$\ptres{\nabla_{T}T}{\nabla_{T}T}=-\frac{(f')^2}{f^2}+f^2g_2\Big(\frac{
c }{f^2}\nn,\frac{ c }{f^2}\nn\Big) = \frac{ c ^2-(f')^2}{f^2}$.
In order to obtain a Frenet curve, we must assume $ c
^2-(f')^2\neq 0$ everywhere. We define
\begin{equation}\label{deltas}
\Delta :=  c ^2-(f')^2, \qquad \delta:=\mathrm{sign}(\Delta)=\pm
1, \qquad \kappa := \frac{\sqrt{\delta\Delta}}{f}.
\end{equation}
According to previous computations and notation, we have
$$
\nabla_TT=\delta \frac{\raiz}{f}\Big(\frac{f'\delta}{\raiz},
\frac{\delta c \nn}{f\raiz}\Big).
$$
Therefore, we can choose
$$N=\Big(\frac{f'\delta}{\raiz}, \frac{\delta c \nn}{f\raiz}\Big),  \qquad
B=\Big(\frac{ c }{\raiz}, \frac{f'\nn}{f\raiz}\Big).
$$
In fact, it is easy to check that $\{T,N,B\}$ is an orthormal
basis along $\alpha$, with $\epsilon_2=\delta=-\epsilon_3$. Let us
check that it is also positive. We recall that we should compare
it with $\{\partial_t,\gamma',\nn\}$, bearing in mind
(\ref{deltas}); that is to say, we compute
$$\det(T,N,B)=\left\vert\begin{array}{ccc} 0 & \frac{f'\delta}{\raiz} & \frac{ c }{\raiz} \\
1/f & 0 & 0 \\ 0 & \frac{\delta c }{f\raiz} & \frac{f'}{f\raiz}
\end{array}
\right\vert =\frac{-1}{f^2\Delta} \left\vert\begin{array}{cc} f' &
c  \\  c  & f'
\end{array}
\right\vert = \frac{1}{f^2}>0.
$$
With this curve $\alpha$, we construct an immersion $\phi$ as in
(\ref{immersion}). By (\ref{e2}), we can study the causal
character of the mean curvature vector $\vec{H}_{\phi}$ of $\phi$.
Thus, we have
$$
\epsilon_2\left(\kappa+\frac{f'}{f}\overline{g}_3\Big(\partial_t,
N\Big)\right)^2 +\epsilon_3\left(\frac{f'}{f}
\overline{g}_3\Big(\partial_t,B\Big)\right)^2 = \delta
\left(\frac{\raiz}{f}-\,\frac{\delta (f')^2}{f\raiz}\right)^2
-\delta \frac{(f')^2}{f^2}\left(\frac{c}{\raiz}\right)^2=$$
$$ = \delta \,\frac{(c^2-2(f')^2)^2-(f')^2 c^2}{\delta f^2\Delta}.
$$
This means that $\vec{H}_{\phi}$ is spacelike (resp.
lightlike/zero, timelike) if, and only if,
$$\mathcal{S}:=\delta\Big((c^2-2f'(t_{0})^2)^2-f'(t_{0})^2 c^2\Big)>0\qquad \hbox{(resp.
$=0$,
$<0$).}
$$
For instance, if the chosen level $t_0$ gives rise to a critical
slide $f'(t_0)=0$, the mean curvature $\vec{H}_{\phi}$ is always
spacelike. \cvd

\begin{proposition} There exist infinitely many embedded tori $\phi:{\mathbb S}^{1}\times {\mathbb S}^{1}\rightarrow
\overline{M}^{4}_{1}$ which are trapped, marginally trapped or
untrapped (whenever $f$ is not constant everywhere).
\end{proposition}
{\it Proof.} We resort to previous example. Pick a point $t_0$
such that, say, $f'(t_0)>0$, and consider a curve $\gamma$ with
constant geodesic curvature $c$ (i.e., the curve $\gamma$ is a
small circle of $\mathbb{S}^2(1/2)$). If $c=2f'(t_0)$ holds,
simple computations show $\delta=1$ and $\mathcal{S}=0$. This
means that the embedding $\phi$ is a marginally trapped surface.
On the other hand, if $c>2f'(t_0)$, we have $\mathcal{S}>0$, which
is the spacelike case. Finally, if we take $f'(t_0)<c<2f'(t_0)$,
then the mean curvature vector of $\phi$ is timelike. \cvd

\begin{remark} {\em Previous proposition can be directly obtained from Theorem \ref{th1}, just by considering suitable CMC tori in $M^{3}={\mathbb S}^{3}$.}
\end{remark}

\subsection{Marginally trapped surfaces crossing expanding and collapsing
regions}\label{ss2.5}

Consider the toy model $\overline{M}^{3}_{1}=I\times_{f}{\mathbb
S}^{2}(1/2)$ associated to a closed FLRW spacetime
$\overline{M}^{4}_{1}=I\times_{f}{\mathbb S}^{3}$. Let
$\alpha:J\subset {\mathbb R}\rightarrow \overline{M}^{3}_{1}$,
$s\mapsto (t(s),\alpha_{2}(s))$ be a curve with $\dot{t}=h(t)$; in
particular,
\[
\ddot{t}(s)=h'(t)\dot{t}(s)=h'(t)h(t).
\]
If we impose the vector field
$\dot\alpha=T=h(t)\partial_{t}+\dot{\alpha}_{2}$ to be unitary,
i.e.
\[
\overline{g}_{3}(\dot{\alpha},\dot{\alpha})=-h(t)^{2}+f(t)^{2}g_{2}(\dot{\alpha}_{2},\dot{\alpha}_{2})=1,
\]
we deduce,
\[
g_{2}(\dot{\alpha}_{2},\dot{\alpha}_{2})=\frac{1+h(t)^{2}}{f(t)^{2}}.
\]
Consider the unitary reparametrization $\tilde{\alpha}_{2}$ of
$\alpha_{2}$, i.e.
\[
\dot{\alpha}_{2}=\rho\dot{\tilde{\alpha}}_{2},\quad
\rho=\frac{\sqrt{1+h(t)^{2}}}{f(t)},\quad
g_{2}(\dot{\tilde{\alpha}}_{2},\dot{\tilde{\alpha}}_{2})=1\quad
\left(\Rightarrow
g(\dot{\tilde{\alpha}}_{2},\dot{\tilde{\alpha}}_{2})=f(t)^{2}\right)
\]
Then, $T$ can be rewritten as
\[
T=h(t)\partial_t+\rho\alfadt.
\]
Moreover
\[
\nabla_{\dot{\tilde{\alpha}}_{2}}\dot{\tilde{\alpha}}_{2}=\nabla^{2}_{\dot{\tilde{\alpha}}_{2}}\dot{\tilde{\alpha}}_{2}-\frac{\overline{g}_{3}(\dot{\tilde{\alpha}}_{2},\dot{\tilde{\alpha}}_{2})}{f(t)}{\rm
grad}_{\overline{g}_{3}}(f)=\nabla^{2}_{\dot{\tilde{\alpha}}_{2}}\dot{\tilde{\alpha}}_{2}+f(t)f'(t)\partial_{t}
\]
\[
\nabla_{\dot{\tilde{\alpha}}_{2}}\partial_{t}=\nabla_{\partial_{t}}\dot{\tilde{\alpha}}_{2}=\frac{f'(t)}{f(t)}\dot{\tilde{\alpha}}_{2},
\quad \nabla_{\partial_t}\partial_t = 0.
\]
Therefore, we deduce:
\[
\begin{array}{rl}
\displaystyle\frac{DT(s)}{ds} & \displaystyle
=\frac{D}{ds}\left(\dot{t}(s)\partial_{t}+\rho\alfadt\right)
=\ddot{t}(s)\partial_{t}+\dot{t}(s)\frac{D\partial_{t}}{ds}+\dot{\rho}\alfadt+\rho\frac{D\alfadt}{ds}
\\
& \\
 &\displaystyle
=\ddot{t}(s)\partial_{t}
+\dot{t}(s)^{2}\nabla_{\partial_{t}}\partial_{t}
+\dot{t}(s)\nabla_{\dot{\alpha}_{2}}\partial_{t}
+\dot{\rho}\alfadt +\rho\left(
\dot{t}(s)\nabla_{\partial_{t}}\alfadt
+\nabla_{\alfadt}\alfadt\right)
\\ & \\
 &
=h'(t)h(t)\partial_{t} +h(t)\rho\nabla_{\alfadt}\partial_{t}
+\dot{\rho}\alfadt +h(t)\rho\nabla_{\partial_{t}}\alfadt
+\rho^2\nabla_{\alfadt}\alfadt
\\ & \\
 &
=h'(t)h(t)\partial_{t} +2h(t)\rho\frac{f'(t)}{f(t)}\alfadt
+\dot{\rho}\alfadt +\rho^2\nabla_{\alfadt}\alfadt
\\ & \\
 &
=h'(t)h(t)\partial_{t}
+\left(2h(t)\rho\frac{f'(t)}{f(t)}+\dot{\rho}\right)\alfadt
+\rho^2\left(\nabla^2_{\alfadt}\alfadt +f(t)f'(t)\partial_t\right)
\\ & \\
& =\left(h'(t)h(t)+\rho^2f(t)f'(t)\right)\partial_{t}
+\left(2h(t)\rho\frac{f'(t)}{f(t)}+\dot{\rho}\right)\alfadt
+\rho^2\nabla^2_{\alfadt}\alfadt,
\\
\end{array}
\]
where the partner function of $\alfadt$ becomes
$$\begin{array}{c}
2h(t)\rho\frac{f'(t)}{f(t)}+\dot{\rho} =\displaystyle
2h(t)\rho\frac{f'(t)}{f(t)}
-h(t)\frac{(1+h(t)^2)f'(t)-f(t)h(t)h'(t)} {f(t)^2\sqrt{}1+h(t)^2}
\\ [0.3em] =
\displaystyle\left(h(t)h'(t)+(1+h(t)^2)\frac{f'(t)}{f(t)}\right)\frac{h(t)}{f(t)\sqrt{1+h(t)^2}}.
\end{array}
$$
Next, let $\{\dot{\tilde{\alpha}}_{2},{\cal N}\}$ be some
orthonormal basis of $TS^{2}(1/2)$ along $\tilde{\alpha}_{2}$.
Assume that $\tilde{\alpha}_{2}$ is a circle (non-necessarily
maximum) in $S^{2}(1/2)$. Then,
\[
\nabla^{2}_{\dot{\tilde{\alpha}}_{2}}\dot{\tilde{\alpha}}_{2}=k
{\cal N}.
\]
In conclusion, we deduce
\[
\nabla_{T}T \ =\
\left(hh'+(1+h^{2})\frac{f'}{f}\right)\left(\partial_{t}+\frac{h}{\sqrt{1+h^{2}}f}\dot{\tilde{\alpha}}_{2}\right)+\frac{(1+h^{2})}{f^{2}}k{\cal
N}.
\]
In particular, if we recall that $\nabla_{T}T=\epsilon_{2}\kappa
N$, we have
\[
\begin{array}{rl}
\overline{g}_{3}(\nabla_{T}T,\nabla_{T}T) &
=\left(hh'+(1+h^{2})\frac{f'}{f}
\right)^{2}\left(\frac{-1}{1+h^{2}}\right)+(1+h^{2})^{2}\frac{k^{2}}{f^{2}}
\\ &
=\epsilon_{2}\kappa^{2}.
\end{array}
\]
Therefore, from the last equality in this expression:
\begin{equation}\label{kappa}
\begin{array}{rl}
\kappa^{2} &
=\epsilon_{2}\left(\frac{(1+h^{2})^{2}k^{2}}{f^{2}}-\frac{1}{1+h^{2}}\left(hh'+(1+h^{2})\frac{f'}{f}\right)^{2}\right)
\\ & =\epsilon_{2}(1+h^{2})\left(
\frac{(1+h^{2})k^{2}}{f^{2}}-\left((\ln\sqrt{1+h^{2}})'+(\ln
f)'\right)^{2}\right).
\end{array}
\end{equation}
Next, we are going to compute the vector field $B$. To this aim,
we write it as:
\[
B=\overline{A}\partial_{t}+\overline{B}\dot{\tilde{\alpha}}_{2}+\overline{C}{\cal
N}
\]
By imposing
$\overline{g}_{3}(T,B)=-h\overline{A}+\overline{B}\sqrt{1+h^{2}}f=0$
we deduce
\begin{equation}\label{*}
\overline{B}=\frac{h}{\sqrt{1+h^{2}}f}\overline{A}.
\end{equation}
On the other hand, by imposing $\overline{g}_{3}(N,B)=0$ we have
$\overline{g}_{3}(\nabla_{T}T,B)=0$, and so,
\begin{equation}\label{**}
\left(hh'+(1+h^{2})\frac{f'}{f}\right)\left(-\overline{A}+\frac{hf}{\sqrt{1+h^{2}}}\overline{B}
\right)+(1+h^{2})k\overline{C}=0.
\end{equation}
Taking into account (\ref{*}) in (\ref{**}) we deduce:
\[
0=-\left(\frac{hh'}{1+h^{2}}+\frac{f'}{f}\right)\overline{A}+(1+h^{2})k\overline{C}.
\]
Therefore, if $k\neq 0$ we have
\begin{equation}\label{***}
\overline{C}=\frac{1}{(1+h^{2})^{2}k}\left(hh'+(1+h^{2})\frac{f'}{f}\right)\overline{A}.
\end{equation}
Next, we impose
$g_{3}(B,B)=-\overline{A}^{2}+f^{2}(\overline{B}^{2}+\overline{C}^{2})=\epsilon_{3}$.
Taking into account (\ref{kappa}), (\ref{*}) and (\ref{***}) in
this expression, one deduces:
\[
\begin{array}{rl}
\overline{A}^{2} &
=\frac{\epsilon_{3}}{-\frac{1}{1+h^{2}}+\frac{f^{2}}{(1+h^{2})^{4}k^{2}}\left(hh'+(1+h^{2})\frac{f'}{f}\right)^{2}}
=-\frac{\epsilon_{3}k^{2}(1+h^{2})^{2}}{f^{2}\left(\frac{(1+h^{2})k^{2}}{f^{2}}-((\ln
\sqrt{1+h^{2}})'+(\ln f)')^{2}
\right)}
\\ &
=\frac{k^{2}(1+h^{2})^{3}}{f^{2}\kappa^{2}}.
\end{array}
\]
Next, we impose that the surface $S$ generated by the curve
$\alpha$ is marginally trapped (\ref{e2}):
\[
\begin{array}{rcl}
\left(\kappa+\frac{f'}{f}\overline{g}_{3}(\partial_{t},N)\right)^{2}
& = & \left(\frac{f'}{f}\overline{g}_{3}(\partial_{t},B)
\right)^{2}
\\
\left(\kappa+\frac{f'}{f\epsilon_{2}\kappa}\overline{g}_{3}(\partial_{t},\nabla_{T}T)\right)^{2}
& = & (f'/f)^{2}\overline{A}^{2}
\\
\left(\epsilon_{2}\kappa^{2}-(\ln f)'\left( hh'+(1+h^{2})(\ln
f)'\right)\right)^{2} & = & ((\ln
f)')^{2}\kappa^{2}\overline{A}^{2}.
\end{array}
\]
If we develop both members of the equation, we deduce:
\[
2((\ln f)')^{2}+\left(
3(\ln\sqrt{1+h^{2}})'+\epsilon\frac{\sqrt{1+h^{2}}k}{f}\right)(\ln
f)'+((\ln\sqrt{1+h^{2}})')^{2}-\frac{(1+h^{2})k^{2}}{f^{2}}=0,
\]
which is a second order equation for $(\ln f)'$. The discriminant
of this equation is:
\[
D=\left((\ln\sqrt{1+h^{2}})'+\frac{3\epsilon
\sqrt{1+h^{2}}k}{f}\right)^{2}.
\]
Therefore, the solutions are
\[
(\ln
f)'_{\pm}=\frac{-3(\ln\sqrt{1+h^{2}})'-\frac{\epsilon\sqrt{1+h^{2}}k}{f}
\pm\left((\ln\sqrt{1+h^{2}})'+3\epsilon\frac{\sqrt{1+h^{2}}k}{f}\right)}{4},
\]
that is,
\[
\left\{\begin{array}{l}
f'_{+}=-\frac{1}{2}(\ln\sqrt{1+h^{2}})'f_{+}+\frac{\epsilon}{2}\sqrt{1+h^{2}}k
\\
f'_{-}=-(\ln\sqrt{1+h^{2}})'f_{-}-\epsilon\sqrt{1+h^{2}}k.
\end{array}\right.
\]
The solution $(\ln f)'_{-}$ implies $\kappa=0$ (recall
(\ref{kappa})). So, take the solution $(\ln f)'_{+}$ with
$\epsilon=+1$. Take also $t(s)=\sin(s)$, and thus,
$h(t)=\pm\sqrt{1-t^{2}}$. In this case, the warping function $f$
must satisfy the differential equation
\begin{equation}\label{g}
f'(t)=\frac{t}{2(2-t^{2})}f(t)+\frac{\sqrt{2-t^{2}}}{2}k,\quad
f(0)=f_{0}>0.
\end{equation}
So, according to previous construction, we will obtain a
marginally trapped surface crossing expanding and collapsing
regions along a closed FLRW spacetime with, say, $I=(-1.3,1.3)$,
if, in addition, the following two properties hold for $f_{0}$ big
enough:
\begin{itemize}
\item[(i)] $f(t)>0$ for all $t\in [-1.3,1.3]$, and

\item[(ii)] $f'(t)$ changes its sign along $(-1,1)$.
\end{itemize}
For (i), first we are going to prove that $f(t)>0$ on $[0,1.3]$.
By contradiction, we assume there exists $t_{1}\in [0,1.3]$ such
that $f(t_{1})\leq 0$. Since $f'(0)=k>0$, $f$ has to start to
decrease at a certain point, and therefore there exists $t_{0}\in
(0,t_{1})$ such that $f'(t_{0})=0$, $f(t_{0})>0$. This is in
contradiction to the fact that
\[
f'(t_{0})=\frac{t_{0}}{2(2-t_{0}^{2})}f(t_{0})+\frac{\sqrt{2-t_{0}^{2}}}{2}k>0.
\]
Next, we focus on $[-1.3,0]$. Let $g(t)$ be the solution of the
problem:
\[
g'(t)=\frac{\sqrt{2-t^{2}}}{2}k,\quad g(0)=f_{0}.
\]
It is straightforward to check that
$$g(t)=\frac{1}{4}kt\sqrt{2-t^{2}}+\frac{k}{2}\arcsin(t/\sqrt{2})+f_{0}\qquad\hbox{on}\;\; [-1.3,0].$$
In particular, we notice that $g(t)>0$ on $[-1.3,0]$ whenever
$f_{0}$ is big enough. We are going to show that, under these
conditions, $g(t)\leq f(t)$ on $[-1.3,0]$. To this aim, define
\[
\Gamma=\{t\in [-1.3,0]: g(t)\leq f(t)\}.
\]
Since $g(0)=f_{0}=f(0)$, it is $0\in \Gamma\neq\emptyset$.
Moreover, $\Gamma=(g-f)^{-1}((-\infty,0])$ is closed in
$[-1.3,0]$. In order to show that $\Gamma=[-1.3,0]$, it suffices
to prove that $\Gamma$ is open, or, equivalently, if $[t,0]\subset
\Gamma$ then $[t-\epsilon,0]\subset\Gamma$ for some $\epsilon>0$.
So, assume that $[t,0]\subset\Gamma$. Then, $f(t)\geq g(t)$, and
thus,
\[
f'(\overline{t})-g'(\overline{t})=\frac{\overline{t}}{2(2-\overline{t}^{2})}f(\overline{t})<0,\quad\hbox{for
all}\;\; \overline{t}\in [t,0].
\]
Therefore,
\[
\int_{t}^{0}f'(\overline{t})d\overline{t}<\int_{t}^{0}g'(\overline{t})d\overline{t},\quad\hbox{and
thus},\quad f(t)>g(t).
\]
By continuity, there exists $\epsilon>0$ such that
$f(\overline{t})\geq g(\overline{t})$ for all $\overline{t}\in
[t-\epsilon,0]$, and so, $[t-\epsilon,0]\subset \Gamma$.
Summarizing, we have proved that property (i) above holds whenever
$f_{0}$ is big enough.

For property (ii), observe that $f'(0)>0$. Moreover, $f(-1)\geq
g(-1)\nearrow\infty$ if $f_{0}\nearrow\infty$. Therefore,
\[
f'(-1)=-\frac{f(-1)}{2}+\frac{k}{2}<0\quad \hbox{if}\quad
f_{0}\;\;\hbox{is big enough}.
\]
Hence, property (ii) also holds whenever $f_{0}$ is big enough.

In conclusion, we have proved the existence of a closed FLRW
spacetime admitting a marginally trapped surface crossing
expanding and collapsing regions.
\begin{remark} {\em Two important subtleties have been omitted in previous
development:
\begin{itemize}
\item[1.] Notice that our approach breaks down at the points where
function $h$ becomes zero, i.e. for $t(s)=\pm 1$; and so, we
cannot ensure, a priori, that our surface is marginally trapped at
the corresponding points. This difficulty is overcome just by
noting that the continuity of the length of the mean curvature
vector, joined to the fact that this length is zero at the rest of
the points, ensures that it is also zero here. \item[2.] A similar
argument shows that our surface is marginally trapped at the
points where $f'$ vanishes (which must exist by property (ii)).
From (\ref{g}) it is straightforward to check that these points
are isolated in $I$, and so, again a continuity argument on the
length of the mean curvature vector ensures that the surface is
also marginally trapped there.
\end{itemize}
}
\end{remark}

%

\section{Marginally trapped tubes in FLRW spacetimes}\label{s3}


In the present paper we propose the following definition of
marginally trapped tube (compare with the definition of MOTT in
\cite{AMS}):
\begin{definition} A smooth manifold $\mathcal{G}$ which admits a foliation $\{\mathcal{S}_{\lambda}: \in \lambda\in\Lambda\}$, is a {\em marginally trapped tube} in a spacetime $\mathcal{M}$ if there is a smooth immersion of codimension 1, $\Phi:\mathcal{G}\rightarrow \mathcal{M}$, such that:
\begin{description}
 \item[\textrm{(A)}] Each $\Phi(\mathcal{S}_{\lambda})$ ($\lambda\in\Lambda$) is a marginally trapped surface in $\mathcal{M}$, and
\item[\textrm{(B)}] $\Phi(\mathcal{S}_{\lambda})\cap \Phi(\mathcal{S}_{\mu})=\emptyset$ for any $\mu,\lambda\in\Lambda$, $\mu\neq \lambda$.
\end{description}
\end{definition}
The second condition is required in order to avoid
self-intersections in the direction of propagation of the tube.

With this definition in mind, we are going to apply our approach
to obtain some information about marginally trapped tubes in FLRW
spacetimes.
We begin with the following direct consequences of the
corresponding existence/non-existence results for marginally
trapped surfaces in Section \ref{s1} (Cor. \ref{c1}, \ref{c2}):
\begin{corollary}\label{c3} (Existence result).
There exist marginally trapped tubes whose $t$-sections are formed
by closed marginally trapped surfaces of any genus in closed
($M^{3}={\mathbb S}^{3}$) FLRW spacetimes.
\end{corollary}

\begin{corollary}\label{c4} (Non-existence result).
Let $\overline{M}_{1}^{4}=I\times_{f} M^{3}$ be a FLRW spacetime
with fiber $M^{3}={\mathbb H}^{3}$. There are no marginally
trapped tubes, with $t$-sections formed by closed marginally
trapped surfaces, crossing $t_{0}$-slices with $|f'(t_{0})|\leq
1$.
\end{corollary}

%
%
%

Next, we are going to give examples of marginally trapped tubes
with any type of causality in closed FLRW spacetimes.

\subsection{Examples of marginally trapped tubes with different
causality}\label{ss3.2}


Given a closed FLRW spacetime
$\overline{M}^{4}_{1}=I\times_{f}\mathbb{S}^{3}$, first we are
going to construct a marginally trapped tube foliated by Clifford
tori (see Section \ref{s1}) and defined for any time.

We define the smooth function
$$h:I\rightarrow (0,\pi/2),\quad h(t)
=\frac{1}{2}\mathrm{arccot}\left(\frac{f'(t)}{2}\right)
=\frac{\pi}{4}-\frac{1}{2}\arctan\left(\frac{f'(t)}{2}\right),$$
whose derivative is
$$ h'(t)=\frac{-f''(t)}{4+f'(t)^2}.
$$
Next, we define the embedding
\[
\phi : I\times \mathbb{S}^1\times\mathbb{S}^1\rightarrow
\overline{M}_1^4, \]
\[\phi(t,e^{i\theta},e^{i\nu})=\left(t, e^{i\theta}\cos(h(t)), e^{i\nu}\sin(h(t))
\right).
\]
We notice that for each $t\in I$, the surface
$\phi(t,-,-):\mathbb{S}^1\times\mathbb{S}^1\rightarrow
\overline{M}_1^4$, is a Clifford torus embedded in the $t$-slice.
By comparing with the expression of the Clifford torus, we
see that the length of the mean curvature of the torus (see (\ref{ec:cm-toro-clifford})) at $t$ is
$\|\vec{H}_u\|$, with $u=h(t)$. A straightforward computation
shows
$$\|\vec{H}_u\|_{u=h(t)} = |2\cot(2u)\mid_{u=h(t)}|=|f'(t)|.
$$
By Theorem \ref{th1}, for each $t$, the torus is a marginally
trapped surface.

Next, we pay attention to the causal character of the embedding
$\phi$. In this sense, we are going to compute the first
fundamental form induced by $\bar{g}_4$ on our surface. To this
aim, first we compute the derivatives
\begin{eqnarray*}
\phi_t & = &  \left(1,-e^{i\theta}\sin(h(t))h'(t),e^{i\nu}\cos(h(t))h'(t) \right), \\
\phi_{\theta} & = & \left(0,i e^{i\theta}\cos(h(t)),0 \right), \\
\phi_{\nu} & = &  \left(0,0,i e^{i\nu}\sin(h(t))\right).
\end{eqnarray*}
A straightforward computation shows
$$\phi^{*}\bar{g}_4 \equiv
\left(
\begin{array}{ccc}
\left(f(t)h'(t)\right)^2-1 & 0 & 0 \\
0 &   \left(f(t)\cos(h(t))\right)^2 & 0 \\
0 & 0 & \left(f(t)\sin(h(t))\right)^2
\end{array}
\right).
$$
Since $0<h(t)<\pi/2$ for any $t\in I$, the derivatives
$\phi_{\theta}$ and $\phi_{\nu}$ are always spacelike. Thus,
everything depends on the derivative $\phi_t$. By recalling the
expressions of $\phi_t$ and $h'(t)$, we obtain
\begin{equation}z(t):=\bar{g}_4(\phi_t,\phi_t) =
-1+\left(\frac{f(t)f''(t)}{4+f'(t)^2}\right)^2. \label{gbarra4}
\end{equation}
This expression can take any value, positive, negative or zero,
depending only on the warping function $f$. We show a list of
particular cases:
\begin{enumerate}
\item We choose non-negative real constants
 $a$ and $b$ such that $a^2=4+b^2$. In particular, $a>b$, which makes the function $f:I=\mathbb{R}\rightarrow(0,\infty)$, $f(t)=a \cosh(t)+b\sinh(t)$ well-defined. A simple computation shows
$z(t)=0$. Therefore, $\phi_t$ is everywhere lightlike, and so is
the corresponding marginally trapped tube. \item We choose real
constants $c_1, c_2>0$. Then, the function
$f:\mathbb{R}\rightarrow(0,\infty)$,
$f(t)=\frac{4+c_1^2}{4c_2}t^2+c_1t+c_2$ is well-defined. A simple
computation shows $z(t)=-3/4$. This implies that $\phi_t$ is
everywhere timelike, and so is the corresponding marginally
trapped tube. \item We define the function
$f:(-1,1)\rightarrow(0,\infty)$, $f(t)=\frac{2}{1-t^2}$. Now, we
see
$$z(t)+1=\left(\frac{f(t)f''(t)}{4+f'(t)^2}\right)^{2}= \frac{6t^2+2}{t^8-4t^6+6t^4+1}.
$$
If we show that $z(t)\geq 1$, then we obtain that $\phi_t$ is
always spacelike, and thus, the marginally trapped tube is also
spacelike. By simple computations, we have that for any $t\in
(-1,1)$, $z(t)+1\geq 2 \Longleftrightarrow 0\geq
t^{2}(t^6-4t^4+6t^2-3)$. Standard computations give that the only
real roots of the equality are $t=0,\pm 1$. This readily proves
$z(t)\geq 1$ for any $t\in(-1,1)$. \item We define the function
$f:\mathbb{R}\rightarrow(0,\infty)$, $f(t)=3+\cos(2 t)$. A
straightforward computation gives
$$ z(t)=-1+\left(\frac{f(t)f''(t)}{4+f'(t)^2}\right)^2=-1+\frac{4\cos^2(2t)(3+\cos(2t))^2}
{(3-\cos(4t))^2}.
$$
It is easy to check $z(0)=15$ and $z(\pi/4)=-1$. Therefore, the
marginally trapped tube changes its causal character with time.
\end{enumerate}

\section{Marginally trapped surfaces in $t$-slices of twisted
spaces}\label{s4}

Assume that the warping function $f$ also depends on the fiber
$M^{3}$, i.e. $f:I\times M^{3}\rightarrow (0,\infty)$. Denote
again by $\overline{M}^{4}_{1}=I\times_{f}M^{3}$ the Lorentzian
twisted manifold given by the product manifold $I\times M^{3}$
endowed with metric $\overline{g}_{4}=-dt^{2}+f^{2}g_{3}$. As in
Section \ref{s1}, let $\varphi:S\rightarrow M^{3}$ be an immersion
of $S$ in $M^{3}$, $\psi:M^{3}\rightarrow I\times_{f}M^{3}$ the
embedding of $M^{3}$ in $I\times_{f}M^{3}$ and $\phi:S\rightarrow
I\times_{f}M^{3}$ the corresponding immersion of $S$ in the
twisted product, both in a $t$-slice. Again from \cite[p.
79]{Chen}, we have the following relation between the
corresponding second fundamental forms:
\[
h_{\phi}(X,Y)=h_{\varphi}(X,Y)+h_{\psi}(X,Y),\quad\hbox{whereby}\;\;
X,Y\in\mathfrak{X}(S).
\]
From \cite[Prop. 2]{PR}, the second fundamental form $h_{\psi}$ is
$$h_{\psi}(X,Y)=-\overline{g}_{4}(X,Y)\frac{{\rm
grad}_{\overline{g}_{4}}f}{f}=\overline{g}_{4}(X,Y)\left(
\frac{\partial_{t}f}{f}\partial_{t}-\frac{1}{f^{3}}{\rm
grad}_{g_{3}} f\right),$$ where we have used that
$${\rm
grad}_{\overline{g}_{4}}f\left(=\sum_{ij}\overline{g}_{4}^{ij}\frac{\partial
f}{\partial x^{i}}\frac{\partial}{\partial
x^{j}}\right)=-\partial_{t} f\partial_{t}+\frac{1}{f^{2}}{\rm
grad}_{g_{3}}f.$$ Hence,
$$h_{\phi}(X,Y)=h_{\varphi}(X,Y)+\overline{g}_{4}(X,Y)\left(
\frac{\partial_{t}f}{f}\partial_{t}-\frac{1}{f^{3}}{\rm
grad}_{g_{3}} f\right).$$ Taking one half of the trace of the
above expression, using an orthonormal frame $\{\partial_{t},
\{E_{i}\}_{i=1}^{3}\}$ w.r.t. the metric $\overline{g}_{4}$, i.e.
$E_{i}=\frac{e_{i}}{f}$ whereby $\{e_{i}\}_{i=1}^{3}$ is the
corresponding orthonormal frame w.r.t. the metric $g_{3}$ on
$M^{3}$ (and $S$), one obtains
\begin{equation}\label{e1'}
\vec{H}_{\phi}=\frac{1}{f^{2}}\left(\vec{H}_{\varphi}-\frac{1}{f}{\rm
grad}_{g_{3}}f\right)+\frac{\partial_{t}f}{f}\partial_{t},
\end{equation}
where $\vec{H}_{\phi}$ and $\vec{H}_{\varphi}$ stand for the mean
curvature associated to $h_{\phi}$ and $h_{\varphi}$,
respectively.
\begin{theorem}\label{th1'} A surface $S$ contained in a $t_{0}$-slice
of twisted $\overline{M}_{1}^{4}=I\times_{f}M^{3}$ is marginally
trapped iff its mean curvature vector satisfies:
\[
\left\|\vec{H}_{\varphi}(\cdot)-\frac{1}{f(t_{0},\cdot)}{\rm
grad}_{g_{3}}f(t_{0},\cdot)\right\|=|\partial_{t}f(t_{0},\cdot)|.
\]
On the other hand, $S$ is trapped iff
\[
\left\|\vec{H}_{\varphi}(\cdot)-\frac{1}{f(t_{0},\cdot)}{\rm
grad}_{g_{3}}f(t_{0},\cdot)\right\|<|\partial_{t}f(t_{0},\cdot)|.
\]
\end{theorem}

\subsection*{Acknowledgments}

The authors are partially supported by the Spanish MEC-FEDER Grant
MTM2007-60731 and by the Grant P09-FQM-4496 (Consejería de
Innovación J. Andalucía with FEDER funds). Also, the second author
is partially supported by the Research Foundation - Flanders
project G.07.0432.

The authors thank Professors Marc Mars and J.M.M. Senovilla for
their useful comments and suggestions. JLF also thanks the
Department of Mathematics, Katholieke Universiteit Leuven, where
part of this work started, for its kind hospitality during his
stay there.


\end{document}